\documentclass[seceq,supplement]{ptptex}
\usepackage{graphicx}
\usepackage{bbm}

\newcommand{\dd}{\mathrm{d}}
\newcommand{\td}[2]{\frac{\dd #1}{\dd #2}}
\newcommand{\pd}[2]{\frac{\partial #1}{\partial #2}}
\newcommand{\fd}[2]{\frac{\delta #1}{\delta #2}}
\newcommand{\mean}[1]{\langle #1 \rangle}
\newcommand{\Int}[1]{\int\dd #1\;}
\newcommand{\IInt}[3]{\int_{#2}^{#3}\dd #1\;}

\newcommand{\ket}[1]{|#1 \rangle}
\newcommand{\bra}[1]{\langle #1|}
\newcommand{\sca}[2]{\langle #1|#2 \rangle}

\renewcommand{\vec}[1]{\mathbf #1}

\DeclareMathOperator{\erfc}{erfc}

\newcommand{\al}{\alpha}

\newcommand{\lam}{\lambda}
\newcommand{\gam}{\gamma}

\newcommand{\sig}{\sigma}
\newcommand{\ra}{\rightarrow}

\newcommand{\id}{\mathbbm 1}
\newcommand{\x}{\vec r}
\newcommand{\nois}{\boldsymbol\xi}
\newcommand{\dsm}{\dot s_\mathrm{med}}
\newcommand{\dst}{\dot s_\mathrm{tot}}
\newcommand{\sm}{s_\mathrm{med}}
\newcommand{\st}{s_\mathrm{tot}}
\newcommand{\hk}{_\mathrm{hk}}  
\newcommand{\ex}{_\mathrm{ex}}  
\newcommand{\Ba}{B^\mathrm{(a)}}
\newcommand{\Be}{B^\mathrm{(e)}}
\newcommand{\Bp}{B^\mathrm{(p)}}
\newcommand{\eq}{^\mathrm{(eq)}}
\newcommand{\tobs}{t_\mathrm{obs}}
\newcommand{\La}{\mathcal L}

\title{Driven Soft Matter:\\ Entropy Production and the
  Fluctuation-Dissipation Theorem}

\author{
  Thomas \textsc{Speck}
}

\inst{
  Department of Chemistry, University of California, Berkeley,
  \\
  Chemical Sciences Division, Lawrence Berkeley National Laboratory, \\
  Berkeley, California 94720, USA
}

\abst{Entropy and the fluctuation-dissipation theorem are at the heart of
  statistical mechanics near equilibrium. Driving a system beyond the linear
  response regime leads to (i) the breakdown of the fluctuation-dissipation
  theorem and (ii) a nonzero entropy production rate. We show how both
  phenomena are related using the general framework of stochastic
  thermodynamics suitable for soft matter systems governed by stochastic
  dynamics and driven through nonconservative forces or external flows. In
  particular, the excess of the fluctuation-dissipation theorem in a
  nonequilibrium steady state compared to equilibrium is related to total
  entropy production. Alternative recent derivations of generalized
  fluctuation-dissipation theorems are sketched and related to each other. The
  theory is illustrated for two systems: a driven single colloidal particle
  and systems driven through simple shear flow.}

\begin{document}

\maketitle


\section{Introduction}

The fluctuation-dissipation theorem (FDT) is one of the cornerstones of
statistical mechanics. Going beyond equilibrium, it connects the response of a
system perturbed slightly out of equilibrium with correlations in
equilibrium~\cite{kubo}. Specifically for a system at temperature $T$, and
Boltzmann's constant set to unity, the FDT reads
\begin{equation}
  \label{eq:fdt:eq}
  TR\eq_{A,h}(t-t') \equiv \left.\fd{\mean{A(t)}}{h(t')}\right|_{h=0}
  = -\partial_t\mean{A(t)B\eq(t')}_0,
\end{equation}
i.e., the response of the system after some perturbation $h$ has been applied
is measured through a time-dependent change of the mean value of $A$. This
response is equal to the time-derivative of a correlation function in
equilibrium involving $A$ and another observable $B\eq$. This second
observable at earlier time is not arbitrary but rather is the conjugate of the
field $h$ with respect to the system's energy $U$, i.e., applying the field
changes the energy as $U\mapsto U-hB\eq$. Onsager's regression principle casts
this remarkable symmetry into words: the decay of spontaneous fluctuations
cannot be distinguished from the decay of a forced fluctuation.

The values of $h$ for which the form~(\ref{eq:fdt:eq}) of the FDT is valid
determine the \textit{linear response regime}. Concepts like linear
irreversible thermodynamics~\cite{groot} and local equilibrium assumptions
have been used successfully to extend certain thermodynamic concepts into this
near-equilibrium regime based on the notion of entropy production. However, a
general theory for going beyond the linear response regime is still missing.
In practical terms, the absence of a concept comparable in universality to the
Gibbs-Boltzmann distribution is probably the biggest obstacle in formulating
and applying a general nonequilibrium thermodynamics. During the last 10-15
years substantial progress has been made on another front through the
formulation and study of nonequilibrium fluctuation relations valid
arbitrarily far from equilibrium. These relations constrain the probability
distributions of quantities like work, heat, and entropy production. The
arguably most famous representatives are the nonequilibrium work relations due
to Jarzynksi~\cite{jarz97,jarz97a} and Crooks~\cite{croo99,croo00} and the
fluctuation theorems for entropy
production~\cite{evan93,gall95,evan02,seif05a}. Subsequently, it has been
shown that thermodynamics can be formulated consistently for driven systems on
the level of single trajectories~\cite{bust05,rito07,seif08}.

The systems of interest to this work are soft matter systems such as colloidal
suspensions, single colloidal particles, biomolecules such as DNA and RNA, as
well as motor proteins such as F$_1$-ATPase. These systems share the property
that they are immersed into a host fluid of well defined temperature and that
they intrinsically operate in nonequilibrium; they are either driven by
chemical gradients, mechanical forces, or external flows. However, we will
require the fluctuations arising from the bath--system interactions to be
described by equilibrium fluctuations. This cannot be strictly true, for an
explicit calculation see, e.g., Ref.~\cite{szam04}. However, several
experimental tests for single colloidal particles~\cite{blic06,spec07,blic07}
confirm the theoretical predictions for measured probability distributions,
thus supporting the validity of this approximation.

The FDT is one of the ubiquitous tools in statistical mechanics and
computational physics. Due to its importance, possible extensions into the
realm of nonequilibrium, especially for glassy dynamics, of been studied for a
long time leading to a number of reviews (a selection is
Refs.~\cite{cris03,cala05,marc08}, see also references therein). In this
paper, based on Ref.~\cite{seif09}, we mainly discuss the connection between
entropy production as defined in the framework of stochastic thermodynamics
and the fluctuation-dissipation theorem. The main ingredient is that, although
the system is driven beyond the \textit{linear
  response regime}, a \textit{linear response} in reaction to a slight
perturbation of the driven system out of its nonequilibrium steady state can
still be defined. Hence, while the equilibrium fluctuation-dissipation theorem
no longer holds in the form~(\ref{eq:fdt:eq}), it can be extended to
nonequilibrium steady states by an additive correction. The principle of
conjugate observables with respect to energy is replaced by conjugate
observables with respect to entropy production. We discuss the relation to
other recent work on the FDT and finally we give specific expressions for the
case of a single colloidal particle and colloidal suspensions or polymers
driven by simple shear flow.


\section{Driven Soft Matter}

In the following, we consider stochastic systems obeying Markovian
dynamics. The state space might either be discrete or continuous, and $x$
denotes a single element in this space. For time-independent forces or rates,
the system will eventually settle in a steady state with probability
distribution $\psi_0(x;\{\lam_j\})$. It is important to note that this steady
state depends on parameters $\lam_j$ that we control externally. To simplify
notations, however, we will often not write this dependence explicitly. The
values of these parameters determine whether the steady state is equilibrium
or a nonequilibrium steady state characterized by a non-vanishing mean entropy
production rate.

\subsection{Continuous state space}

Soft matter systems such as colloidal particles or biomolecules immersed in a
fluid are often well described by overdamped Langevin dynamics. The fluid acts
as the heat reservoir, i.e., we assume that the fluid remains in equilibrium
with temperature $T$ even though we drive the immersed subsystem. In addition,
the fluid might have an imposed flow profile $\vec u(\x)$. We consider the
system to be composed of $N$ `units' (colloidal particles or monomers) with
positions $\x_k$, ignoring internal degrees of freedom. The configuration of
the system is then given by $x\equiv\{\x_1,\dots,\x_N\}$. The time-evolution
of the probability distribution $\psi(x,t)$ is governed by the Smoluchowski
equation
\begin{equation}
  \label{eq:sm}
  \partial_t\psi + \sum_{k=1}^N \nabla_k\cdot(\vec v_k\psi) = 0,
\end{equation}
where
\begin{equation}
  \label{eq:lmv}
  \vec v_k = \vec u(\x_k) + \mu_0\vec F_k = \vec u(\x_k) +
  \mu_0\left[-\nabla_kU + \vec f_k - T\nabla_k\ln\psi \right]
\end{equation}
is the local mean velocity and $\mu_0$ is the bare mobility due to
friction. Any deviation of the local mean velocity $\vec v_k$ from the imposed
flow profile $\vec u(\x)$ has to be caused by a force $\vec F_k$ exerted on
the $k$-th particle. We allow for three contributions to this force: (i)
conservative forces $-\nabla_kU$ due to the potential energy $U(x)$, (ii)
nonconservative forces $\vec f_k$, and (iii) ``thermodynamic'' forces arising
from the stochastic interactions between system and the surrounding fluid. In
equilibrium, i.e., in the absence of external flows and nonconservative
forces, detailed balance holds which amounts to $\vec v_k=0$. The
thermodynamic forces ensure that the equilibrium Gibbs-Boltzmann
distribution $\psi_0\propto e^{-U/T}$ is the solution of
Eq.~(\ref{eq:sm}). Although a force on a particle in principle leads to a
distortion of the flow and the coupling of forces, here we will ignore
hydrodynamic interactions between particles.

An equivalent dynamic prescription on the level of single stochastic
trajectories is the Langevin equation
\begin{equation}
  \label{eq:lang}
  {\dot\x}_k = \mu_0\left[-\nabla_kU + \vec f_k \right] + \nois_k,
\end{equation}
where the noise $\nois_k$ explicitly models the stochastic interactions
between particles and the surrounding fluid. The noise has zero mean and
correlations
$\mean{\nois_k(t)\nois^T_{k'}(t')}=2\mu_0T\id\delta_{kk'}\delta(t-t')$.

\subsection{Stochastic thermodynamics}

Stochastic thermodynamics is a conceptual framework combining energetics along
single trajectories with the definition of a stochastic
entropy~\cite{seif08}. We first extend the idea by
Sekimoto~\cite{seki97,seki98} to define work and heat along single stochastic
trajectories to the situation where external flows are
present~\cite{spec08}. In a thermodynamic context, the work is the change in
energy that is controlled externally. The work rate reads~\cite{spec08}
\begin{equation}
  \label{eq:work}
  \dot w = \sum_j\dot\lam_j\pd{U}{\lam_j}
  + \sum_{k=1}^N\vec u(\x_k)\cdot\nabla_kU
  + \sum_{k=1}^N\vec f_k\cdot[\dot\x_k-\vec u(\x_k)].
\end{equation}
The first term is the work spent to change the potential energy through
changing control parameters. The other two terms are due to the external flow
and the nonconservative forces, respectively. Conservation of energy in form
of the first law of thermodynamics then leads to the heat rate
\begin{equation}
  \label{eq:heat}
  \dot q = \dot w - \td{U}{t}
  = \sum_{k=1}^N [-\nabla_kU+\vec f_k]\cdot[\dot\x_k-\vec u(\x_k)].
\end{equation}
Hence, the total force times displacement of the particles equals the heat
dissipated into the surrounding fluid.

\begin{figure}[t]
  \centering
  \includegraphics[width=.7\linewidth]{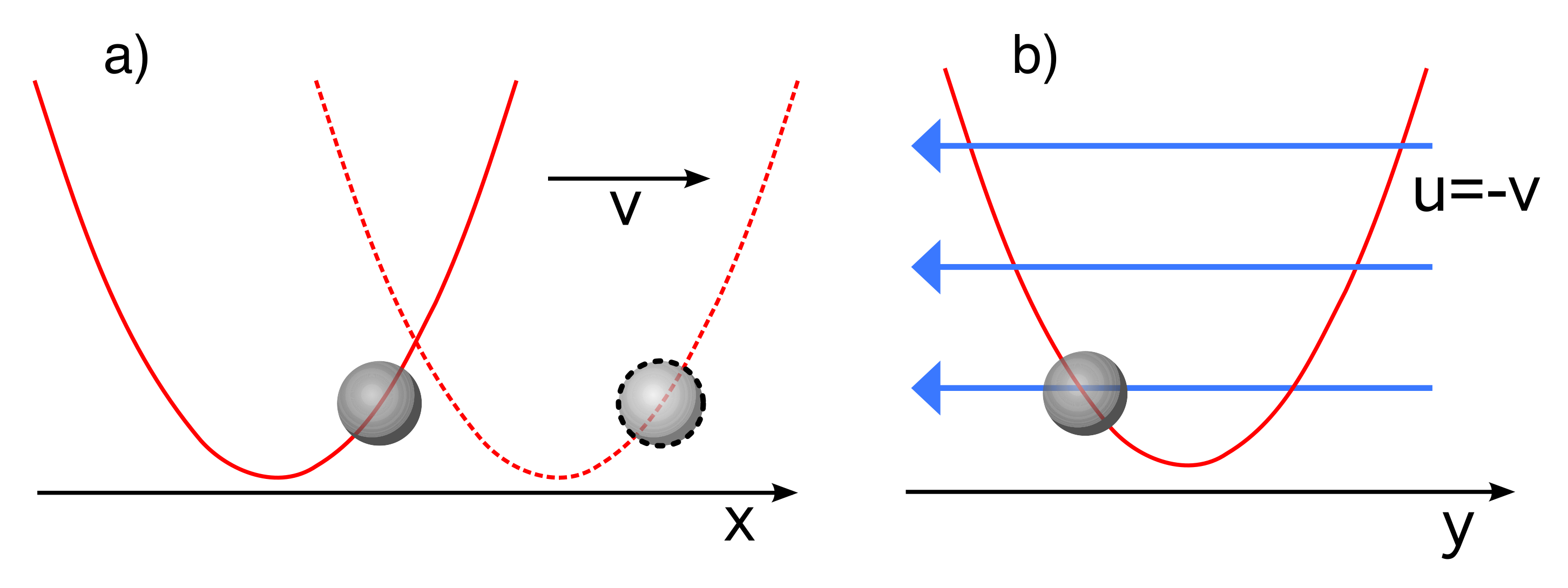}
  \caption{Colloidal particle dragged by optical tweezers through a viscous
    fluid. a) Laboratory frame of reference. b) Co-moving frame of reference:
    an observer moving such that the trap appears stationary would see a flow
    trying to advect the particle. The work is invariant in both scenarios.}
  \label{fig:frame}
\end{figure}

To make these ideas transparent and as an example, consider the prototypical
colloidal particle dragged by optical tweezers, see Fig.~\ref{fig:frame}. The
optical trap is modeled as a harmonic potential with strength $k$ and focus
position $\lam$. This system has been studied comprehensively both
theoretically~\cite{zon03,spec05} and experimentally~\cite{trep04}. The
parameter we control externally is the position of the trap
center. Alternatively, one could control the trap strength. The potential
energy reads $U(x;\lam)=(k/2)(x-\lam)^2$ and the work rate following
Eq.~(\ref{eq:work}) is $\dot w=\dot\lam k(x-\lam)$, where $x$ is the position
of the particle in the laboratory frame. An observer moving with (constant)
velocity $v=\dot\lam$ measures the particle position $y=x-vt$. As expected,
the expression for the work rate, $\dot w=-kyv$, is invariant under such a
change of the frame of reference although $\dot\lam=0$ since now the fluid
appears to flow with velocity $u=-v$. The corresponding expressions for the
heat are $\dot q=-k(x-\lam)\dot x=-k(v+\dot y)$.

The final step is the introduction of a stochastic entropy defined
as~\cite{seif05a,seif08}
\begin{equation}
  \label{eq:ent:def}
  s(t) \equiv -\ln\psi(x(t),t;\{\lam_j(t)\}).
\end{equation}
Averaging this expression leads to the well known Shannon entropy
\begin{equation}
  \label{eq:ent:shannon}
  \mean{s(t)} = -\Int{x}\psi(x,t)\ln\psi(x,t).
\end{equation}
Taking the time-derivative of Eq.~(\ref{eq:ent:def}), we obtain the balance
equation
\begin{equation}
  \label{eq:ent:eom}
  \td{s}{t} = -\dsm + \dst,
\end{equation}
i.e., the total entropy production rate $\dst$ is the sum of the change of
entropy of the system $\dd s/\dd t$ and the change of medium entropy
$\dsm$. Since we demand the fluid to stay in equilibrium, we can identify
the dissipated heat with the change of fluid entropy through the Clausius
relation $\dsm=\dot q/T$.

\subsection{Discrete state space}

Discrete states might be intrinsic but often they arise through some spatial
or temporal coarse graining procedure. A biochemically motivated example for
the former is the discrete motion of F$_1$-ATPase~\cite{yasu98}. Transitions
between states occur with rates $W(x\ra y)$. Analogous to the continuous case
discussed in the previous section, the thermodynamic notions of work and heat
can be defined consistently~\cite{schm06}. However, stochastic entropy itself
might be introduced in a more abstract way regardless of whether the system of
interest is coupled to a heat bath~\cite{seif05a}. A single stochastic
trajectory $x(t)$ of length $\tobs$ consists of $K$ jumps at times
$0<t_\al<\tobs$ from state $x_{\al-1}$ to state $x_\al$ with a given initial
state $x_0$. The total change of stochastic entropy along this trajectory (in
a steady state) can be written
\begin{equation}
  \label{eq:ent:dis}
  \begin{split}
    \Delta s \equiv s(\tobs)-s(0)
    &= -\sum_{\al=1}^K \ln\frac{\psi_0(x_\al)}{\psi_0(x_{\al-1})} \\
    &= \sum_{\al=1}^K \ln\frac{W(x_\al\ra x_{\al-1})}{W(x_{\al-1}\ra x_\al)}
    - \sum_{\al=1}^K \ln\frac{\psi_0(x_\al)W(x_\al\ra x_{\al-1})}
    {\psi_0(x_{\al-1})W(x_{\al-1}\ra x_\al)} \\
    &\equiv -\sm[x(t)] + \st[x(t)],
  \end{split}
\end{equation}
where in the second line we have expanded the sum by inserting the transition
rates. The sums run over all transitions. Following Eq.~(\ref{eq:ent:eom}), we
interpret the first sum as the change of medium entropy with rate
\begin{equation}
  \label{eq:11}
  \dsm(t) = -\sum_{\al=1}^K \delta(t-t_\al)
  \ln\frac{W(x_\al\ra x_{\al-1})}{W(x_{\al-1}\ra x_\al)}.
\end{equation}
The second term in Eq.~(\ref{eq:ent:dis}) is the total entropy production. In
equilibrium, detailed balance holds with $\psi_0(x)W(x\ra y)=\psi_0(y)W(y\ra
x)$. Hence, in equilibrium the total entropy production becomes identically
zero. Moreover, $\mean{\st}\geqslant0$, where the equal sign holds in
equilibrium. This identification of entropy and the corresponding fluctuation
relations for their probability distributions have been illustrated
experimentally for a driven single defect in diamond~\cite{schu05,tiet06}.


\section{The Fluctuation-Dissipation Theorem}

Before discussing the FDT in nonequilibrium, we will briefly introduce an
abstract notation particularly apt to treat continuous and discrete state
space on equal footing and to see the general structure of the FDT we want to
unveil. Using the bra-ket notation, the system is described by the state
vector $\ket{\psi(t)}$ from which the probability distribution is read off as
$\psi(x,t)=\sca{x}{\psi(t)}$. We define a reference state $\bra{-}$ through
$\sca{-}{\psi(t)}=1$ which expresses the conservation of probability. The time
evolution of the state vector is given through
\begin{equation}
  \label{eq:master}
  \partial_t\ket{\psi(t)} = \hat L\ket{\psi(t)}
\end{equation}
with an operator $\hat L$ obeying $\bra{-}\hat L=0$. The steady state
$\ket{\psi_0}$ corresponds to the right eigenvector, $\hat L\ket{\psi_0}=0$,
of this time evolution operators with eigenvalue zero. In this picture,
the observables correspond to operators with expectation values
\begin{equation}
  \label{eq:mean}
  \mean{A(t)} \equiv \bra{-}\hat A\ket{\psi(t)}.
\end{equation}
For diagonal observables, $\bra{x}\hat A\ket{x'}=A(x)\delta(x-x')$. This
expression reduces to the familiar $\mean{A(t)}=\Int{x}A(x)\psi(x,t)$ upon
inserting the identity $\id=\Int{x}\ket{x}\bra{x}$. Similar expressions where
the integration is replaced by a sum are obtained for discrete state
spaces. We denote the average value of $A$ in the steady state by
$\mean{A}_0$. Finally, two-point correlations of two observables $A$ and $B$
in the steady state become
\begin{equation}
  \label{eq:corr}
  \mean{A(t)B(0)}_0 \equiv \bra{-}\hat A e^{t\hat L}\hat B\ket{\psi_0}.
\end{equation}
For completeness, we give two explicit expressions for $\hat L$. First, the
matrix elements of the time-evolution operator for the Smoluchowski
equation~(\ref{eq:sm}) read
\begin{equation}
  \label{eq:sm:elem}
  \bra{x}\hat L\ket{x'} = -\sum_{k=1}^N\nabla_k\cdot\left\{\vec
    u(\x_k)+\mu_0[-\nabla_kU+\vec f_k-T\nabla_k] \right\}\delta(x-x').
\end{equation}
Second, for a general Markov process on a discrete state space, the
time-evolution operator is the usual left stochastic matrix with elements
\begin{equation}
  \label{eq:5}
  \bra{x}\hat L\ket{x'} = W(x'\ra x) - \delta_{xx'}\sum_{y\neq x} W(x\ra y).
\end{equation}

\subsection{Formal derivation}

Let $\lam$ be one control parameter, and without loss of generality let
$\lam=0$ for the steady state. Changing $\lam$ will perturb the system away
from its steady state with the following evolution governed by
Eq.~(\ref{eq:master}). For small $\lam$, the system will response linearly in
the sense
\begin{equation}
  \label{eq:resp:mean}
  \mean{A(t)} = \mean{A}_0 + \IInt{t'}{-\infty}{t}R_{A,\lam}(t-t')\lam(t')
  + \mathcal O(\lam^2),
\end{equation}
i.e., the mean at later times is a linear functional of the perturbation
$\lam(t)$. The response function $R_{A,\lam}(t)$ connecting both is formally
given through
\begin{equation}
  \label{eq:resp}
  R_{A,\lam}(t-t') \equiv
  \left.\fd{\mean{A(t)}}{\lam(t')}\right|_{\lam=0} \qquad (t'\leqslant t).
\end{equation}
Due to causality, the response function vanishes for $t'>t$.

The calculation of the response function~(\ref{eq:resp}) for general Markovian
processes is well known. For completeness and later reference, we repeat the
derivation here, following the route of Agarwal~\cite{agar72} and H\"anggi and
Thomas~\cite{hang82}. The formal solution of Eq.~(\ref{eq:master}) for a, due
to the perturbation, time-dependent $\hat L(t)$ reads \begin{equation}
  \label{eq:3}
  \ket{\psi(t)} = e_+^{\IInt{t'}{-\infty}{t}\hat L(t')}\ket{\psi_0}
\end{equation}
given that the system at $t\rightarrow-\infty$ has been prepared in the steady
state. The exponential is to be understood in the time-ordered sense. We then
split the time evolution operator, $\hat L(t)=\hat L+\lam(t)\delta\hat L$, and
use the expression~(\ref{eq:mean}) to obtain
\begin{equation}
  \label{eq:fdt:gen}
  R_{A,\lam}(t-t') = \bra{-}\hat Ae^{(t-t')\hat L}\delta\hat L\ket{\psi_0}
  = \mean{A(t)B(t')}_0
\end{equation}
after performing the functional derivative with respect to $\lam(t')$. Hence,
the response function can be expressed as a correlation function of the
observable $A$ with the operator $\delta\hat L$. Although formally correct, it
is of little practical use as long as we do not find a 'physical'
representation $B$ of this observable in the sense that it can be expressed in
terms of, in principle, measurable quantities. For a first representation,
we assume the observable $\delta\hat L$ to be diagonal, $\delta\hat
L=\Int{x}\ket{x}\Ba(x)\bra{x}$, with
\begin{equation}
  \label{eq:Ba}
  \Ba(x) \equiv \frac{\bra{x}\delta\hat L\ket{\psi_0}}{\sca{x}{\psi_0}}.
\end{equation}
This result indeed has been known for more than thirty years and we call it
the 'Agarwal' form. However, $\Ba$ still seems to have no transparent physical
meaning. Moreover, for complex systems with a large number of degrees of
freedom, the explicit form of the stationary distribution $\psi_0(x)$ in
general is neither available experimentally nor from numerical simulations.

In the next subsections we will discuss two more representations of $B$
labeled by different superscripts $B^{(i)}$. To this end it is crucial to
realize that a whole class of representations for the observable $B$ exist
that all lead to the same FDT~(\ref{eq:fdt:gen}). The formal reason is that we
have some freedom in expressing the state $\delta\hat L\ket{\psi_0}$. Beyond
the representations discussed below, there are in principle infinitely many
variants of the FDT since with $B^{(1)}\cong B^{(2)}$, where $\cong$ denotes
the equivalence of observables, any normalized linear combination
\begin{equation}
  \label{eq:1}
  \frac{1}{c_1+c_2}\left[c_1 B^{(1)}+c_2B^{(2)}\right]
\end{equation}
with $c_{1,2}$ real will be admissible.

\subsection{Role of stochastic entropy}

We want to establish the connection between the stochastic
entropy~(\ref{eq:ent:def}) and the FDT~\cite{seif09}. We consider two steady
states separated by a small $\lam$ with state vectors $\ket{\psi_0}$ and
$\ket{\psi_0}+\lam\ket{\delta\psi}$, respectively. Using that these state
vectors are the right eigenvectors of the corresponding evolution operators
with eigenvalue zero,
\begin{equation}
  \label{eq:6}
  \hat L\ket{\delta\psi} = -\delta\hat L\ket{\psi_0}
\end{equation}
holds to first order in $\lam$. Inserting this into Eq.~(\ref{eq:fdt:gen}), we
obtain
\begin{equation}
  \label{eq:fdt:e}
  R_{A,\lam}(t-t') = \pd{}{t'}\bra{-}\hat Ae^{(t-t')\hat
    L}\ket{\delta\psi} = \mean{A(t)\Be(t')}_0
\end{equation}
which is independent of $\delta\hat L$. The physical observable is now given
as
\begin{equation}
  \label{eq:Be}
  \Be(x,\dot x) = \td{}{t}\frac{\delta\psi(x)}{\psi_0(x)}
  = -\left.\pd{}{\lam}\td{s}{t}\right|_{\lam=0}.
\end{equation}
The total time derivative in the first expression is along a single trajectory
starting at $x$. One should keep in mind that this expression is to be
averaged over trajectories in the correlation function~(\ref{eq:2}). Expanding
Eq.~(\ref{eq:ent:def}) up to first order in $\lam$ leads to
$-\delta\psi(x)/\psi_0(x)$ and, finally, we interchange the order of
derivations to obtain $\Be$. In contrast to Eq.~(\ref{eq:Ba}), $\Be$ in
general depends also on $\dot x$ and hence is a non-diagonal observable.

The conceptual advantage of Eq.~(\ref{eq:Be}) is that it leads to an
observable which is conjugate to the perturbation parameter with respect to
entropy production, in the same spirit that the observable $B\eq$ is conjugate
with respect to energy in equilibrium. To recover the equilibrium
form~(\ref{eq:fdt:eq}) of the FDT, we employ the stationary distribution in
equilibrium given by the Gibbs-Boltzmann distribution
\begin{equation}
  \label{eq:gb}
  \psi_0(x;\lam) = e^{-[U(x;\lam)-\mathcal F(\lam)]/T},
\end{equation}
where $\mathcal F$ is the free energy. The system entropy simply becomes
$Ts(t)=U(x(t))-\mathcal F$. Eq.~(\ref{eq:fdt:e}) then reads
\begin{equation}
  \label{eq:2}
  TR\eq_{A,\lam}(t-t') = \pd{}{t'}\mean{A(t)[-\partial_\lam U]_{\lam=0}(t')}_0
  + \pd{}{t'}\left[\mean{A}_0\td{\mathcal F}{\lam}\right].
\end{equation}
The free energy is constant along trajectories. The term in the square
brackets is then constant and the second term vanishes. The expression
$\partial_\lam s=(1/T)\partial_\lam U$ appears in the FDT, which acquires the
well-known form~(\ref{eq:fdt:eq}). Hence, we can identify
$B\eq(x)=-\left.\partial_\lam U(x;\lam)\right|_{\lam=0}$ as expected.

Considering the general case of a nonequilibrium steady state, the
observable~(\ref{eq:Be}) can be used to find a link to the equilibrium
case~(\ref{eq:fdt:eq}). Using Eq.~(\ref{eq:ent:eom}), we split
\begin{equation}
  \label{eq:Be:split}
  \Be = \partial_\lam\dsm - \partial_\lam\dst
\end{equation}
into two terms. In Ref.~\cite{seif09} it has been shown that the second term
vanishes in equilibrium. Hence, in equilibrium the
FDT~(\ref{eq:fdt:gen}) can then also be written in the form
\begin{equation}
  \label{eq:ent:eq}
  R\eq_{A,\lam}(t-t') = \mean{A(t)[\partial_\lam\dsm](t')}_0.
\end{equation}
Now consider the same system driven into a nonequilibrium steady state by a
driving force corresponding to the parameter $\lam$. Perturbing the system by
a small change of the same driving force will leave $\partial_\lam\dsm$
unaltered. We can thus keep the correlation function on the right hand side of
Eq.~(\ref{eq:ent:eq}), now evaluated under nonequilibrium conditions, and
subtract the second term in Eq.~(\ref{eq:Be:split}) that involves the
observable conjugate to total entropy production to obtain
\begin{equation}
  \label{eq:10}
  R_{A,\lam}(t-t') = \mean{A(t)[\partial_\lam\dsm](t')}_0 -
  \mean{A(t)[\partial_\lam\dst](t')}_0.
\end{equation}
The second correlation function quantifies the excess compared to equilibrium.
Such a splitting has been mentioned before~\cite{cris03,diez05} without
recognizing the meaning of the excess term as a correlation with the total
entropy production. Harada and Sasa have discussed the connection of this
excess with energy dissipation~\cite{hara05}.

\subsection{Path weight approach}

An alternative way to write the average~(\ref{eq:mean}) is to use the path
integral formalism leading to 
\begin{equation}
  \label{eq:mean:path}
  \mean{A(t)} = \Int{x} A(x) \int^{x(t)=x} [\dd x(t)]\; P[x(t);\lam(t)],
\end{equation}
where the path integral runs over all trajectories which end in $x$ at time
$t$. The weight of a single trajectory is $P[x(t)]=e^{-S[x(t)]}$ with
stochastic action $S$. Calculating the response function~(\ref{eq:resp}) by
taking the functional derivative, we obtain
\begin{equation}
  \label{eq:13}
  R_{A,\lam}(t-t') = \Int{x\dd x'} A(x) \int_{x(t')=x'}^{x(t)=x}
  [\dd x(t)]\; P_0[x(t)] B^{(\mathrm{p})}(x',\dot x')\psi_0(x'),
\end{equation}
where $P_0$ is the path weight in the steady state and the path integral now
sums over all paths starting in $x'$ at earlier time $t'$ and ending in $x$ at
time $t$. We therefore again find the FDT in the general
form~(\ref{eq:fdt:gen}) but now with the observable representation
\begin{equation}
  \label{eq:12}
  \Bp(x,\dot x) = -\left.\fd{S[x(t);\lam(t)]}{\lam(t')}\right|_{\lam=0}
  = -\partial_\lam \La(x,\dot x).
\end{equation}
For the last equality, we have used that the action can be written as
$S=\Int{t}\La$. The explicit expressions read\footnote{There is a Jakobian
  involved in the change of variables from $\nois$ to $x$ which we dropped
  since it does not contribute to $-\partial_\lam\La$.}
\begin{equation}
  \label{eq:7}
  \La(t) = \frac{1}{4\mu_0T}\sum_{k=1}^N [\nois_k(t)]^2
\end{equation}
with $\nois_k$ given through the Langevin equation~(\ref{eq:lang}) and
\begin{equation}
  \label{eq:14}
  \La(t) = \sum_y W(x(t)\ra y) - \sum_{\al=1}^K\delta(t-t_\al)
  \ln W(x_{\al-1}\ra x_\al)
\end{equation}
for continuous and discrete state space, respectively.

A similar approach has been used by Baiesi \textit{et al.} to derive yet
another form of the FDT by relating the path weight of the perturbed process
with the stationary path weight~\cite{baie09}. However, certain forms of this
FDT have been know for a long time~\cite{cugl94,cala05}. For example,
realizing that a force perturbation is equivalent to a perturbation of the
noise leads immediately to
\begin{equation}
  \label{eq:8}
  TR_{A,\vec f_k}(t-t') = \mu_0\left\langle\fd{A(t)}{\nois_k(t')}\right\rangle
  = \frac{1}{2}\mean{A(t)\nois_k(t')}.
\end{equation}
This has been exploited in Ref.~\cite{spec06}.

\subsection{Hatano-Sasa relation approach}

The term steady state thermodynamics has been coined for a phenomenological
theory~\cite{oono98,sasa06} promoting the splitting of the total dissipated
heat into a housekeeping heat $q\hk$ and an excess heat $q\ex$. We introduce a
pseudo-potential $\phi(x)$ via the steady state probability \begin{equation}
  \label{eq:phi}
  \psi_0(x) = e^{-\phi(x)}, \qquad \phi(x) \neq U(x)/T
\end{equation}
in analogy to, but different from, the Gibbs-Boltzmann distribution. The
excess heat is $q\ex/T=Y-\Delta\phi$, where $Y$ is the transition functional
defined as
\begin{equation}
  \label{eq:Y}
  Y[x(t)] \equiv \Int{t} \dot\lam_i(t)\pd{\phi(x(t);\{\lam(t)\})}{\lam_i}.
\end{equation}
For this transition functional, the fluctuation relation
\begin{equation}
  \label{eq:hs}
  \mean{e^{-Y}} = 1
\end{equation}
holds~\cite{hata01}. For completeness, note that also the housekeeping heat
$q\hk\equiv q-q\ex$ fulfills a fluctuation relation~\cite{spec05a}.

We define the variables
\begin{equation}
  \label{eq:4}
  X_j \equiv -\pd{\phi}{\lam_j}
\end{equation}
conjugate to the parameter $\lam_j$ with respect to the pseudo-potential
$\phi$. Expanding the Hatano-Sasa relation~(\ref{eq:hs}) in powers of the
$\lam$'s, Prost \textit{et al.}~\cite{pros09} derived the FDT
\begin{equation}
  \label{eq:fdt:hs}
  \mean{X_j} = \IInt{t'}{-\infty}{t}R_{jk}(t-t')\lam_k(t'), \qquad
  R_{jk}(t-t') = -\partial_t\mean{X_j(t)X_k(t')}_0,
\end{equation}
where $\mean{X_j}_0=0$ by construction. Noting that in a steady state the
stochastic entropy~(\ref{eq:ent:def}) becomes $s(t)=\phi(x(t))$, we see that
this result is equivalent to the FDT~(\ref{eq:fdt:e}). However,
Eq.~(\ref{eq:fdt:e}) seems to be more general since it does not restrict the
observable that measures the response.


\section{Illustrations}
\label{sec:illu}

For an illustration, we consider two systems that have been discussed in
detail previously~\cite{spec09,seif09}: sheared soft matter systems and a
single particle moving in a periodic potential.

\subsection{Shear driven systems}

The system consists of $N$ particles with positions
$x\equiv\{\x_1,\dots,\x_n\}$. The FDT in such shear driven suspensions has
been addressed numerically~\cite{barr00,bert02} and in the framework of
mode-coupling theory~\cite{szam04,krug09}. Invariant quantities~\cite{baul08}
constitute an exact result for systems driven through boundaries with an
otherwise unaltered Hamiltonian. In contrast, in this example the system is
driven through an imposed flow with profile $\vec u(\x)=(\gam y,0,0)^T$, where
$\gam$ is the strain rate. Shearing the fluid builds up stress with
off-diagonal element 
\begin{equation}
  \label{eq:15}
  \sig_{xy}(x) \equiv \sum_{k=1}^N y_k\pd{U}{x_k},
\end{equation}
which will be the crucial quantity in the following.

The time-evolution operator is given by Eq.~(\ref{eq:sm:elem}). Perturbing the
strain rate $\lam\mapsto\gam$, we obtain $\delta
L=-\sum_{k=1}^Ny_k\pd{}{x_k}$. The first representation of the conjugate
observable is the diagonal 'Agarwal' form~(\ref{eq:Ba})
\begin{equation}
  \label{eq:9}
  T\Ba(x) = T\frac{\delta L\psi_0(x)}{\psi_0(x)} 
  = \sig_{xy}(x) - \bar\sig_{xy}(x).
\end{equation}
The second term here is
\begin{equation}
  \label{eq:16}
  \bar\sig_{xy}(x) \equiv \sum_{k=1}^N [U(x)-T\phi(x)] 
  = -\frac{1}{\mu_0} \sum_{k=1}^N y_k(\nu_{k,x}-\gam y_k),
\end{equation}
which involves the $x$-component of the local mean
velocity~(\ref{eq:lmv}). The medium entropy production rate following
Eq.~(\ref{eq:heat}) is
\begin{equation}
  \label{eq:18}
  T\dsm = \sum_{k=1}^N [\dot\x_k-\vec u(\x_k)]\cdot(\nabla_k U)
\end{equation}
and therefore $T\partial_\gam\dsm=\sig_{xy}$. The representation $\Be$ based
on the splitting of the entropy production~(\ref{eq:Be:split}) is therefore
also diagonal and has to coincide with the first form, $\Be=\Ba$. A different
representation is obtained as
\begin{equation}
  \label{eq:19}
  T\Bp = \sum_k \frac{1}{2\mu_0}\left(\dot x_k+\mu_0\pd{U}{x_k}
    -\gam y_k\right)y_k
  = \frac{1}{2}\sig_{xy} + \frac{1}{2\mu_0} \sum_k y_k(\dot x_k-\gam y_k)
\end{equation}
following the path integral approach~(\ref{eq:12}) with
Eq.~(\ref{eq:7}). Combining the two different forms, we arrive at
\begin{equation}
  \label{eq:17}
  T[2\Bp-\Ba] = \frac{1}{\mu_0}\sum_{k=1}^N y_k(\dot x_k-\nu_{k,x}).
\end{equation}
This expression also quantifies some kind of stress, where the velocity in
$x$-direction is measured with respect to the flow.

\begin{figure}[t]
  \centering
  \includegraphics[width=.6\linewidth]{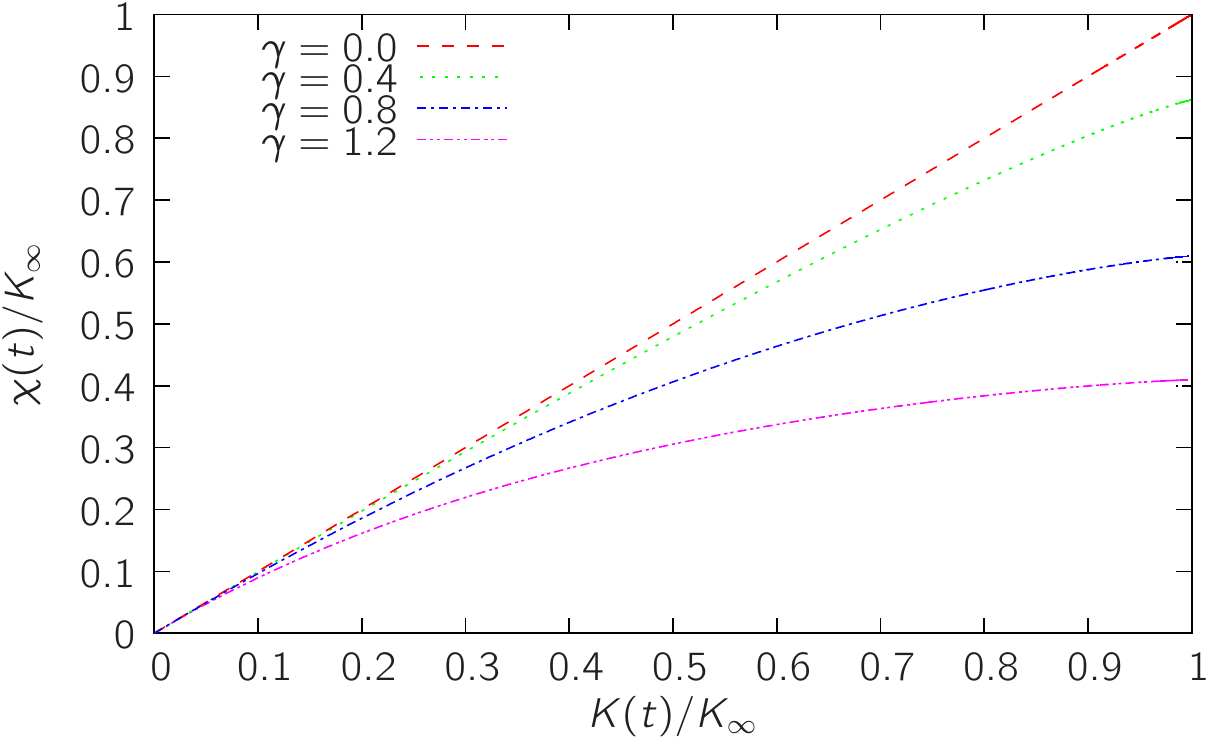}
  \caption{Rouse polymer: The integrated response $\chi(t)$
    [Eq.~(\ref{eq:rouse:chi})] plotted \textit{vs} the integrated correlation
    function $K(t)$ [Eq.~(\ref{eq:rouse:K})] for different strain rates $\gam$
    and normalized by $K_\infty\equiv K(t\ra\infty)$. In equilibrium
    ($\gam=0$), this plot is a straight line with $T\chi(t)=K(t)$.}
  \label{fig:rouse}
\end{figure}

Knowing the different representations the FDT can acquire might help to
measure, but of course does not specify, the actual functions $R(t)$, $C(t)$,
and $I(t)$. Explicit expressions for response and correlation functions have
been obtained for the case of a Rouse polymer~\cite{doi} in
Ref.~\cite{spec09}. The potential energy reads
$U(x)=(k/2)\sum_{k=1}^{N-1}|\x_{k+1}-\x_k|^2$. In the limit $N\ra\infty$, the
following expressions for integrated response and stress auto correlations
have been calculated,
\begin{gather}
  \label{eq:rouse:chi}
  \chi(t) \equiv \IInt{t}{0}{\infty}R_{\sig_{xy},\gam}(t)
  = T\left[(1-e^{-t})+\sqrt{\pi t}\erfc(\sqrt t)\right], \\
  \label{eq:rouse:K}
  K(t) \equiv \IInt{t}{0}{\infty} 
  \left[ \mean{\sig_{xy}(t)\sig_{xy}(0)}-\mean{\sig_{xy}}^2 \right]
  = T^2\left[(1+\gam^2)(1-e^{-t})+\sqrt{\pi t}\erfc(\sqrt t)\right].
\end{gather}
Here, the time $t$ has been scaled by the fundamental relaxation time
$\tau_1\equiv(2\mu_0k)^{-1}$. In Fig.~\ref{fig:rouse}, the integrated response
$\chi(K)$ as function of the integrated correlation is plotted for different
strain rates $\gam$. While for $\gam=0$ we observe a straight line
corresponding to $T\chi(t)=K(t)$, the excess and therefore the deviation from
the straight line increase with increasing strain rate $\gam$.

\subsection{Single particle in a periodic potential}

Another example we have studied previously is that of a driven single
colloidal particle moving in one dimension in a periodic
potential~\cite{spec06,seif09}. The Langevin equation reads
\begin{equation}
  \label{eq:20}
  \dot x = \mu_0[-\partial_xU(x)+f] + \xi \equiv \mu_0F(x) + \xi.
\end{equation}
Here we choose $\lam\mapsto f$, i.e., we perturb the driving force $f$. The
perturbation operator $\delta L=-\mu_0\partial_x$ with Eq.~(\ref{eq:Ba})
produces the diagonal Agarval form $\Ba(x)=\nu(x)-\mu_0F(x)$. Following the
path weight approach leads to $\Bp(x,\dot x)=(1/2)[\dot x-\mu_0F(x)]$. A
straightforward combination of both results amounts to 
\begin{equation}
  \label{eq:21}
  \Be = 2\Bp - \Ba = \dot x - \nu.
\end{equation}
Since form Eq.~(\ref{eq:heat}) $T\partial_f\dsm=\dot x$ follows we can
identify $T\partial_f\dst\cong\nu$, i.e., the excess involves as observable
the local mean velocity~(\ref{eq:lmv}). The idea of identifying the excess
with the local mean velocity has lead to a formulation where the FDT acquires
its equilibrium form through a transformation into the Lagrangian reference
frame moving with the local mean velocity~\cite{chet08,chet09,gome09}.
Fluctuations in this co-moving frame are essentially equilibrium fluctuations
and the driving manifests itself in the mean values of observables. Of course,
this result also crucially depends on fixing the bath fluctuations to be
equilibrium fluctuations.

\section{Summary}

In summary, the general FDT for Markovian dynamics is considered. Following
Ref.~\cite{seif09}, a generalized FDT is derived where the conjugate
observable is determined with respect to \textit{entropy production} of the
system in contrast to equilibrium where the conjugate observable is determined
with respect to \textit{energy}. Alternative ways to derive generalized FDTs
for driven systems are sketched based on the path-integral and on the
Hatano-Sasa relation~(\ref{eq:hs}). It is shown how the results of different
approaches can be combined to produce new conjugate observables through linear
combination. The general theory is illustrated for shear driven systems where
the strain rate is the perturbation parameter and a driven single colloidal
particle with the driving force taking the role as the perturbation parameter.

\section*{Acknowledgments}

The presented work is rooted in my Ph.D. thesis~\cite{speck} and continued in
collaboration with Udo Seifert. Part of this work has been presented at the
workshop 'Frontiers in Nonequilibrium Physics: Fundamental Theory, Glassy \&
Granular Materials, and Computational Physics' as part of the Yukawa
International Seminars (YKIS) in 2009. I would like to thank the organizers,
in particular Hisao Hayakawa and Shin-ichi Sasa. I am deeply grateful to
Valentin Blickle and Clemens Bechinger for fruitful experimental
collaborations, and Jakob Mehl for illuminating discussions on the role of
external flow. I thank David Limmer for a critical reading of the
manuscript. Finally, I acknowledge financial support by Deutsche
Forschungsgemeinschaft and Alexander-von-Humboldt foundation as well as the
Helios Solar Energy Research Center which is supported by the Director, Office
of Science, Office of Basic Energy Sciences of the U.S. Department of Energy
under Contract No.~DE-AC02-05CH11231.



\end{document}